\documentclass[%
 reprint,
 aps,
 pra,
 twocolumn,
 amsmath,amssymb,
 floatfix,
]{revtex4-2}

\usepackage{graphicx}
\usepackage{dcolumn}
\usepackage{bm}
\usepackage{hyperref}
\usepackage{tikz}
\usetikzlibrary{quantikz}
\usepackage{listings}
\usepackage{xcolor}
\usepackage{braket}

\begin{document}

\preprint{APS/123-QED}

\title{QuScope: An Open-Source Python Framework for Quantum-Circuit Simulation of Transmission Electron Microscopy}

\author{Sean D. Lam}
\email{s\_lam2023@coloradocollege.edu}
\affiliation{Department of Physics, Colorado College, Colorado Springs, CO, United States}
\affiliation{Department of Chemistry and Biochemistry, Colorado College, Colorado Springs, CO, United States}

\author{Roberto dos Reis}
\email{roberto.reis@northwestern.edu}
\affiliation{Department of Materials Science and Engineering, Northwestern University, Evanston, IL, United States}
\affiliation{The NUANCE Center, Northwestern University, Evanston, IL, United States}
\affiliation{International Institute of Nanotechnology, Northwestern University, Evanston, IL, United States}

\date{\today}

\begin{abstract}
Image formation in transmission electron microscopy (TEM) is governed by the coherent evolution of the electron wavefunction through the specimen and the objective lens. This physics maps naturally onto the gate model of quantum computation. We present QuScope, an open-source Python framework that expresses the complete TEM image-formation pipeline as quantum circuits. The $N\times N$ electron wavefunction is amplitude-encoded in $2\log_2 N$ qubits, and every optical element, including phase-grating transmission, Fresnel propagation between specimen slices, and the aberrated objective lens, is implemented as a diagonal unitary conjugated by quantum Fourier transforms. On this foundation, QuScope v0.2.0 implements validated imaging pipelines, covering conventional TEM under the phase-object approximation, full multislice CTEM and STEM for thick specimens. All quantum results reported here come from exact, noise-free statevector simulation of the circuits on classical hardware, and every result is validated against a classical twin implementation. Quantum and classical multislice exit waves agree to unit fidelity, and all physical constants are verified against standard references. We provide transpiled quantum-resource estimates for each algorithm, an analysis of the diagonal-synthesis bottleneck that governs near-term hardware execution, and a fully tested, documented, and pip-installable package. QuScope v0.2.0 is available at \url{https://github.com/QuScope/QuScope}.
\end{abstract}


\maketitle

\section{Introduction}
\label{sec:intro}

Transmission electron microscopy (TEM) and its associated techniques are among the most powerful tools for materials characterization at the atomic scale. Interpreting experimental images quantitatively, however, almost always requires forward simulation of the imaging process. An electron wavefunction is propagated through a model specimen and the microscope optics, and the simulated observable is then compared with experiment~\cite{Kirkland2020}. The workhorse algorithm for this task is the multislice method~\cite{CowleyMoodie1957,GoodmanMoodie1974}, which alternates transmission through thin specimen slices with Fresnel propagation between them. Each propagation step is a Fourier-space filter, so a classical simulation of an $N \times N$ wavefunction costs $\mathcal{O}(N^2 \log N)$ per slice. This cost repeats over slices, defoci, probe positions (in scanning TEM, one full multislice per scan pixel), and thermal configurations. For large fields of view, thick specimens, or frozen-phonon ensembles, these costs compound into a substantial computational burden that shapes what is practical in quantitative electron microscopy. Mature classical packages mitigate this cost through compiled or GPU-accelerated implementations and algorithmic approximations, for example the PRISM scattering-matrix method \cite{PRISM_fastSTEM, RANGELDACOSTA2021103141} and abTEM \cite{abtem}. These packages remain fundamentally classical, polynomial-cost methods that operate on intensity-domain simulation.

The structure of this problem is strikingly quantum mechanical. The simulated object is a wavefunction, and the operations applied to it are unitary, for example position-space phase masks, Fourier transforms, and momentum-space phase masks. The observables are intensities, that is, measurement statistics. This makes TEM image formation a natural candidate for expression in the quantum circuit model~\cite{NielsenChuang2010}. An $N \times N$ complex wavefunction can be amplitude-encoded in only $n_q = 2\log_2 N$ qubits, the fast Fourier transform is replaced by the quantum Fourier transform (QFT) acting separably on row and column registers, and every optical element becomes a diagonal unitary. The mapping is exact rather than heuristic. As we demonstrated for CTEM under the phase-object approximation \cite{lam2026quantum}, the quantum circuit implements the same physics as the classical algorithm, with an exponential compression of the state representation.

Here, we present QuScope, an open-source Python framework that realizes this mapping end to end~\cite{quscope2026}. QuScope currently expresses the TEM image-formation pipeline, covering phase-object imaging and multislice propagation for both plane-wave (CTEM) and focused-probe (STEM) illumination, as Qiskit~\cite{qiskit2024} quantum circuits. These circuits run today on statevector simulators and are designed for eventual deployment on quantum hardware. Two design principles distinguish the framework. Every quantum algorithm ships with a classical twin, a closed-form or FFT-based reference implementation against which the quantum results are validated automatically, to unit fidelity where the mapping is exact. The framework also practices honest accounting. We report transpiled gate counts and depths for the actual circuits, and we identify the synthesis cost of arbitrary diagonal unitaries, which grows exponentially with qubit number, as the concrete obstacle separating exponential state compression from practical quantum advantage. We also describe the structured-potential strategies that may overcome it.

QuScope is a physically exact reformulation of electron-optical simulation in the language of quantum circuits, a validated research and teaching platform for quantum-algorithm development in microscopy, and a resource-estimation testbed for the approaching era in which wave-optical simulation may become a genuine quantum-hardware workload. On current noisy intermediate-scale quantum (NISQ) devices~\cite{Preskill2018}, the circuits presented here are executable only at small grid sizes, and we quantify precisely why.

\subsection{Contributions}

This paper makes the following contributions:
\begin{enumerate}
\item A complete, open-source mapping of the TEM image-formation pipeline (phase grating $\rightarrow$ QFT $\rightarrow$ contrast transfer function $\rightarrow$ inverse QFT) onto quantum circuits, under the phase-object approximation (Sec.~\ref{sec:ctem});
\item Quantum multislice simulation for thick specimens, alternating phase-grating and Fresnel-propagation diagonal unitaries (Sec.~\ref{sec:multislice});
\item Quantum STEM in single-slice and multislice form, using a focused, aberrated probe scanned position by position, with high-angle annular dark-field (HAADF), annular dark-field (ADF), annular bright-field (ABF), and bright-field (BF) detector synthesis from the exit-wave momentum distribution (Sec.~\ref{sec:stem});
\item Systematic validation against classical theory and literature constants, enforced by continuous integration (Sec.~\ref{sec:validation}), and transpiled quantum-resource analysis for every algorithm (Sec.~\ref{sec:resources}).
\end{enumerate}

Quantum electron-diffraction modes, thermal-diffuse-scattering channels, and a Bloch-wave eigensolver based on quantum phase estimation are under active development and outlined as roadmap items in Sec.~\ref{sec:outlook}.

\section{Quantum Circuits for Wave Optics}
\label{sec:background}

\subsection{Quantum circuit model}

A quantum circuit acts on a register of qubits with a sequence of unitary gates, followed by measurement. Three phenomena distinguish this computational model from its classical counterpart. The first is \emph{superposition}. A register of $n$ qubits occupies a state
\begin{equation}
\ket{\psi} = \sum_{j=0}^{2^n-1} c_j \ket{j},
\label{eq:qubit-state-expansion}
\end{equation}
a normalized complex vector over all $2^n$ computational basis states simultaneously. The second is \emph{entanglement}. Multi-qubit states generally cannot be factored into independent single-qubit states, which enables correlations with no classical description. The third is \emph{interference}. Unitary evolution manipulates complex amplitudes, so computational paths can cancel or reinforce. This is the mechanism by which quantum algorithms concentrate probability on desired outcomes.

For wave-optical simulation, the relevant observation is more elementary. The state vector of a qubit register is a discretized complex wavefunction, and the gates available in the circuit model include exactly the operations that electron optics requires.

\subsection{Amplitude encoding of the electron wavefunction}
\label{sec:encoding}

QuScope's central object is the discretized wavefunction $\psi(x_a, y_b)$ on an $N \times N$ grid ($N = 2^{n}$). It is stored as the amplitudes of an $n_q$-qubit register, $n_q = 2n$.
\begin{equation}
\ket{\psi} \;=\; \sum_{a=0}^{N-1}\sum_{b=0}^{N-1} \psi(x_a, y_b)\, \ket{a}\otimes\ket{b},
\label{eq:encoding}
\end{equation}
normalized to $\sum_{ab} |\psi_{ab}|^2 = 1$. A $256\times256$ wavefunction thus occupies 16 qubits, and each doubling of the linear grid size costs two additional qubits, an exponential compression of the state representation relative to the $N^2$ complex numbers stored classically. As with the diagonal unitaries of Sec.~\ref{sec:bottleneck}, this compression is not free in general. Preparing an arbitrary amplitude profile costs $\mathcal{O}(2^{n_q})$ gates \cite{Shende2006}. Only structured profiles, such as the plane-wave and aperture-limited probe states used throughout this paper, admit efficient preparation circuits \cite{grover2002}.

All electron-optical elements used in this work reduce to one of two primitives.
\begin{enumerate}
\setlength{\itemsep}{0.25em}
\item \textbf{Diagonal unitaries} implement position- or momentum-space phase masks.
\begin{equation*}
U_D = \mathrm{diag}\!\left(e^{i\phi_0}, \ldots, e^{i\phi_{2^{n_q}-1}}\right).
\end{equation*}
Transmission through a thin specimen slice, the objective-lens aberration function, and the Fresnel propagator are all of this form.
\item \textbf{The two-dimensional QFT}, implemented separably as one $n$-qubit QFT on the row register and one on the column register. It interconverts the position and momentum representations exactly as the FFT does classically, at a cost of $\mathcal{O}(n^2)$ gates per register rather than $\mathcal{O}(N \log N)$ arithmetic operations.
\end{enumerate}
Because both primitives are exactly unitary, the full imaging pipeline built from them reproduces classical wave optics with no approximation beyond the shared discretization.

\subsection{NISQ constraints and the diagonal-synthesis bottleneck}
\label{sec:bottleneck}

Current quantum hardware operates in the NISQ regime~\cite{Preskill2018}, with on the order of $10^2$ qubits, two-qubit gate error rates at the $10^{-3}$--$10^{-2}$ level, and coherence times that bound practical circuit depth. Qubit \emph{count} is not the obstacle for QuScope's circuits. Sixteen qubits suffice for a $256\times256$ wavefunction. The obstacle is gate count. Synthesizing an \emph{arbitrary} $n_q$-qubit diagonal unitary requires $\mathcal{O}(2^{n_q})$ elementary gates~\cite{Shende2006,Welch2014}, that is, one rotation's worth of work per grid pixel. The exponential compression of the state is therefore paid back, in the general case, by an exponential cost of loading arbitrary potentials into phases. Section~\ref{sec:resources} quantifies this trade for the actual transpiled circuits, and Sec.~\ref{sec:outlook} discusses routes around it when the phase function is structured. For example, when it admits a sparse expansion in an appropriate basis (e.g., Walsh/Fourier-like series) or has only a small number of distinct phase values, diagonal circuits can be much smaller than the worst-case scaling~\cite{Welch2014}.

In the present release all simulations execute on Qiskit's exact statevector simulator, with hardware-transpilation utilities provided for resource estimation and small-scale device experiments.

\section{Package Architecture}
\label{sec:architecture}

QuScope is organized so that every quantum algorithm faces a classical reference implementation of the same physics.

The \texttt{quantum\_ctem} package contains the four imaging pipelines described below, together with supporting infrastructure such as amplitude-encoding utilities, a contrast-transfer-function calculator, momentum-space helpers, Kirkland-parametrized material builders (MoS$_2$, graphene), circuit-optimization and hardware-transpilation tools, and IBM Quantum backend wrappers. The \texttt{ctem} and \texttt{simulations} packages provide the classical twins, FFT-based phase-object and multislice simulators built on the same Kirkland scattering-factor tables~\cite{Kirkland2020} that feed the quantum pipelines.

\begin{lstlisting}[
  language={},
  basicstyle=\ttfamily\footnotesize,
  columns=fixed,
  basewidth=0.55em,
  keepspaces=true,
  showstringspaces=false,
  upquote=true,
  breaklines=false,
  frame=single
]
src/quscope/
  quantum_ctem/  quantum pipelines (this paper)
    quantum_ctem_circuit.py        CTEM (POA)
    quantum_multislice_circuit.py  CTEM multislice
    quantum_stem.py                STEM (POA)
    quantum_stem_multislice.py     STEM multislice
    ctf_calculator.py, hamiltonian.py,
    momentum_space.py, materials/, backends/
  ctem/          classical references
    kirkland_potential.py, wpoa_simulator.py,
    multislice_simulator.py
  simulations/   classical multislice/WPO utils
  utils/         Kirkland scattering-factor tables
\end{lstlisting}

Validator classes (\texttt{QuantumClassicalValidator}, \texttt{QuantumClassicalMultisliceValidator}) run both implementations on identical inputs and report state fidelity $|\braket{\psi_q|\psi_c}|^2$ and root-mean-square error. These validators are exercised by a continuous-integration test suite on Python 3.9--3.12, so the quantum--classical agreement documented in Sec.~\ref{sec:validation} is enforced on every commit rather than asserted once. QuScope requires Python $\geq$ 3.9 and Qiskit $\geq$ 2.0, and is distributed under the MIT license.

Every demonstration in this paper is produced through this public API, with no private or paper-only simulation code involved, so each figure doubles as a usage example of the package. Equally important for interpreting the results, \emph{every quantum circuit in this paper is executed with Qiskit's exact statevector simulator on classical hardware}. No figure shows a quantum-device measurement; hardware execution is analyzed prospectively, through the transpiled resource counts and device-connectivity estimates of Sec.~\ref{sec:resources}. Table~\ref{tab:api} maps each figure to the QuScope entry points that generate it; the generating script ships in the repository's paper source.

\begin{table}[tb]
\centering
\caption{Public QuScope entry points behind each demonstration. All classes and functions are importable from \texttt{quscope.quantum\_ctem} unless noted; projected potentials are built with \texttt{quscope.ctem.KirklandPotential}.}
\label{tab:api}
\small
\setlength{\tabcolsep}{4pt}
\begin{tabular}{@{}l p{0.68\linewidth}@{}}
\hline\hline
Figure & QuScope entry points \\
\hline
Fig.~\ref{fig:ctf} & \texttt{CTFCalculator} (\texttt{ctf}, \texttt{calculate\_point\_resolution})\newline probe formation from \texttt{quantum\_stem} \\
Fig.~\ref{fig:ctem} & \texttt{QuantumCTEMCircuit.\allowbreak simulate}\newline \texttt{QuantumClassicalValidator.\allowbreak compare} \\
Fig.~\ref{fig:multislice} & \texttt{QuantumMultisliceCircuit.\allowbreak simulate}\newline \texttt{QuantumClassicalMultislice}\newline \texttt{Validator} \\
Fig.~\ref{fig:propagation} & \texttt{fresnel\_\allowbreak propagator\_\allowbreak phase}\newline per-slice \texttt{DiagonalGate}/\texttt{QFTGate} subcircuits evolved with Qiskit \texttt{Statevector} \\
Fig.~\ref{fig:stem} & \texttt{build\_\allowbreak probe\_\allowbreak circuit}; \texttt{STEMDetectors.\allowbreak masks}\newline consistency vs.\ \texttt{run\_\allowbreak stem}, \texttt{run\_\allowbreak stem\_\allowbreak multislice} \\
\hline\hline
\end{tabular}
\end{table}

\section{Quantum CTEM Imaging Pipeline}
\label{sec:ctem}

\begin{figure*}[tb]
\centering
\includegraphics[width=\textwidth]{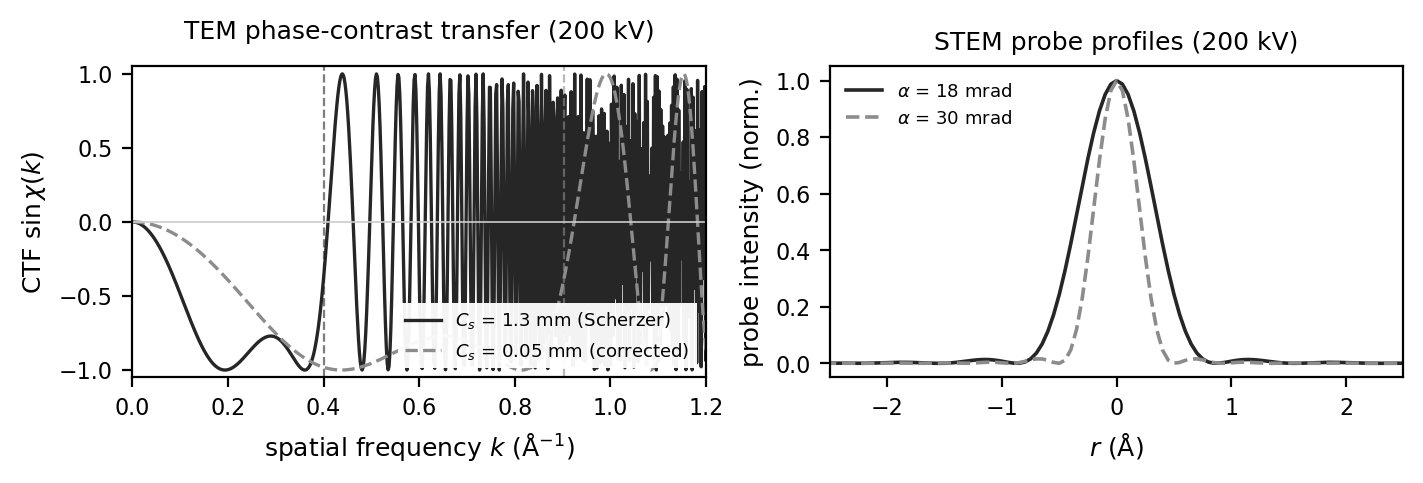}
\caption{Transfer characteristics of the optical settings used in this paper, from QuScope's \texttt{CTFCalculator} at 200~kV. On the left, the TEM phase-contrast transfer function $\sin\chi(k)$ at the extended Scherzer defocus $\Delta\!f = -1.2\sqrt{C_s\lambda}$~\cite{scherzer} is shown for a conventional ($C_s=1.3$~mm, $\Delta\!f=-685.2$~\AA) and an aberration-corrected ($C_s=0.05$~mm, $\Delta\!f=-134.4$~\AA) instrument; vertical lines mark the point resolutions $d = 0.66\,(C_s\lambda^3)^{1/4}$ (2.50 and 1.11~\AA). On the right, aberration-free STEM probe intensity profiles are shown for 18 and 30~mrad convergence semi-angles (FWHM 0.70 and 0.40~\AA).}
\label{fig:ctf}
\end{figure*}

\subsection{Physics}

For a thin specimen the phase-object approximation (POA) reduces specimen interaction to a multiplicative transmission function,
\begin{equation}
t(\mathbf{r}) = e^{\,i\sigma V_z(\mathbf{r})},
\label{eq:wpoa}
\end{equation}
the full phase grating rather than its linearized, weak-phase-object limit $1 + i\sigma V_z(\mathbf{r})$; QuScope implements Eq.~(\ref{eq:wpoa}) exactly, with no small-phase truncation.
where $V_z(\mathbf{r})$ is the projected electrostatic potential (V\,\AA) and $\sigma$ is the relativistic interaction constant~\cite{Kirkland2020}
\begin{equation}
\sigma = \frac{2\pi}{\lambda V}\,\frac{m_0 c^2 + eV}{2 m_0 c^2 + eV},
\label{eq:sigma}
\end{equation}
with $V$ the accelerating voltage and $\lambda$ the relativistic electron wavelength
\begin{equation}
\lambda = \frac{h}{\sqrt{2 m_0 e V \left(1 + \dfrac{eV}{2 m_0 c^2}\right)}}.
\label{eq:lambda}
\end{equation}
At 200~kV, $\lambda = 0.02508$~\AA{} and $\sigma = 0.7288 \times 10^{-3}$~rad\,V$^{-1}$\AA$^{-1}$; QuScope's implementations of Eqs.~(\ref{eq:sigma}) and (\ref{eq:lambda}) reproduce standard reference values to better than $0.1\%$ (Sec.~\ref{sec:validation}).

The objective lens acts in momentum space through the aberration function
\begin{equation}
\chi(k) = \pi \lambda \Delta\!f\, k^2 + \tfrac{1}{2}\pi C_s \lambda^3 k^4,
\label{eq:chi}
\end{equation}
with defocus $\Delta\!f$, spherical aberration $C_s$, and $k$ the spatial frequency in \AA$^{-1}$ (Kirkland's convention; QuScope also supports the fifth-order term $\tfrac{\pi}{3}C_5\lambda^5 k^6$). The recorded image is $I(\mathbf{r}) = |\psi_{\mathrm{image}}(\mathbf{r})|^2$ where $\psi_{\mathrm{image}} = \mathcal{F}^{-1}\!\left[ e^{i\chi(k)}\, \mathcal{F}[t\,\psi_0] \right]$.

Figure~\ref{fig:ctf} characterizes the transfer properties of the two optical settings used throughout this paper, computed with the package's \texttt{CTFCalculator} class. \texttt{ctf(k)} evaluates $\sin\chi(k)$ from a \texttt{CTFParameters} record, and \texttt{calculate\_scherzer\_defocus()} and \texttt{calculate\_point\_resolution()} return the derived optima. For plane-wave TEM, the phase-contrast transfer function at Scherzer defocus \cite{scherzer} gives a point resolution of 2.50~\AA{} for a conventional $C_s = 1.3$~mm instrument and 1.11~\AA{} for an aberration-corrected $C_s = 0.05$~mm instrument. The latter resolves graphene's 1.42~\AA{} bonds in Fig.~\ref{fig:ctem}. For STEM, the same aberration function determines the probe, formed in momentum space by the probe machinery of the \texttt{quantum\_stem} module. The 18 and 30~mrad aperture semi-angles used in Secs.~\ref{sec:multislice} and \ref{sec:stem} produce probes of 0.70 and 0.40~\AA{} full width at half maximum, well matched to resolving the SrTiO$_3$ column spacing.

\subsection{Circuit}

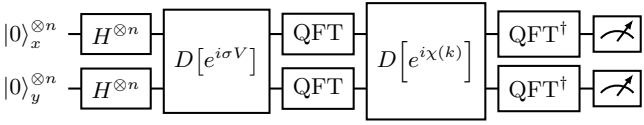
\begin{figure}[tb]
\centering
\resizebox{\columnwidth}{!}{%
\begin{quantikz}[column sep=0.18cm, row sep={0.75cm,between origins}]
\lstick{$\ket{0}^{\otimes n}_{x}$} & \gate{H^{\otimes n}} & \gate[2]{D\!\left[e^{i\sigma V}\right]} & \gate{\mathrm{QFT}} & \gate[2]{D\!\left[e^{i\chi(k)}\right]} & \gate{\mathrm{QFT}^{\dagger}} & \meter{} \\
\lstick{$\ket{0}^{\otimes n}_{y}$} & \gate{H^{\otimes n}} & & \gate{\mathrm{QFT}} & & \gate{\mathrm{QFT}^{\dagger}} & \meter{}
\end{quantikz}}
\caption{Quantum CTEM pipeline [Eq.~(\ref{eq:pipeline})] for an $N\times N$ image with $n = \log_2 N$ qubits per spatial register. The phase grating $D[e^{i\sigma V}]$ and lens transfer function $D[e^{i\chi}]$ are diagonal unitaries; the two-dimensional (I)QFT acts separably on the row and column registers. For multislice simulation (Sec.~\ref{sec:multislice}) the grating--QFT--propagator--IQFT block repeats once per slice before the lens is applied.}
\label{fig:circuit}
\end{figure}

Figure~\ref{fig:circuit} shows the pipeline as implemented by \texttt{QuantumCTEMCircuit}. The register is prepared in the uniform superposition $H^{\otimes n_q}\ket{0}$, the plane wave $\psi_0 = \mathrm{const}$, after which the four optical elements act.
\begin{equation}
\ket{\psi_{\rm image}} = \mathrm{QFT}^{\dagger}\, D[e^{i\chi}]\, \mathrm{QFT}\, D[e^{i\sigma V}]\, H^{\otimes n_q} \ket{0},
\label{eq:pipeline}
\end{equation}
where $D[\cdot]$ denotes the diagonal unitary with the indicated phases sampled on the grid, and QFT is the separable two-dimensional transform of Sec.~\ref{sec:encoding}. Measuring the register in the computational basis samples pixels with probability $|\psi_{\rm image}|^2$, the image intensity, while statevector simulation returns the full complex wavefunction for validation.

\begin{figure*}[tb]
\centering
\includegraphics[width=\textwidth]{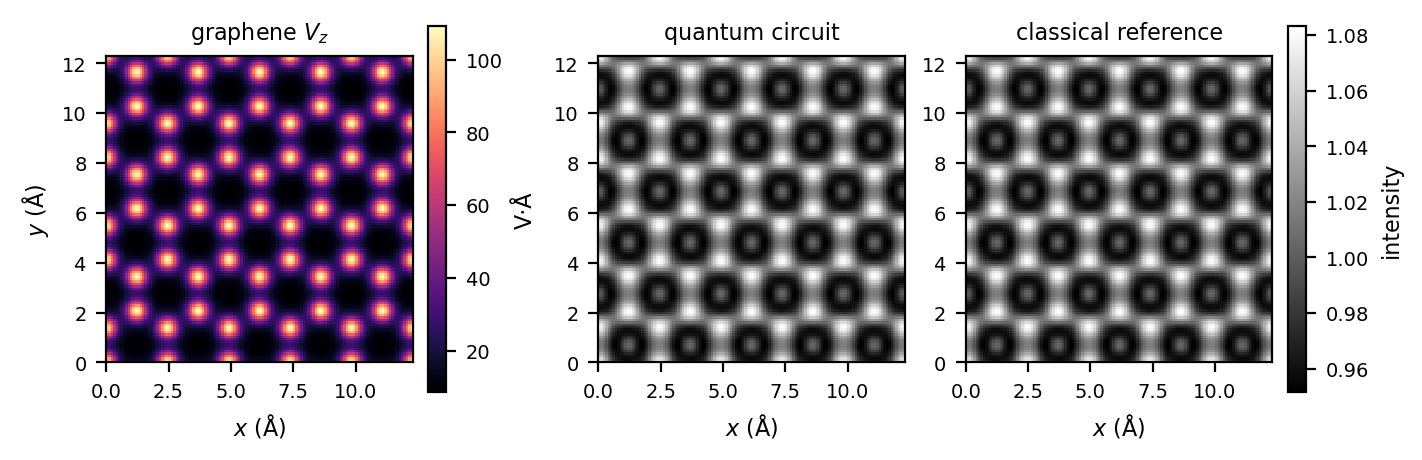}
\caption{Quantum CTEM (POA) of a periodic graphene sheet ($\approx5\times5$ unit cells, 60 atoms) at 200~kV, $128\times128$ grid (14 qubits), aberration-corrected Scherzer conditions ($C_s = 0.05$~mm, $\Delta\!f = -134$~\AA). From left to right, the panels show the projected potential (Kirkland parametrization with Debye--Waller smearing; C--C bond 1.42~\AA), the image intensity from the quantum circuit [Eq.~(\ref{eq:pipeline})] showing the classic dark-atom honeycomb contrast, and the classical FFT reference. Quantum and classical fidelity $|\braket{\psi_q|\psi_c}|^2 = 1.000000$. Exact statevector simulation.}
\label{fig:ctem}
\end{figure*}

Figure~\ref{fig:ctem} shows the pipeline applied to a periodic graphene sheet of $\approx 5\times5$ unit cells (60 atoms, 12.3~\AA{} field of view, $128\times128$ grid at 0.096~\AA/pixel, a 16\,384-amplitude wavefunction on 14 qubits) at 200~kV under aberration-corrected Scherzer conditions ($C_s = 0.05$~mm, $\Delta\!f = -134$~\AA) so the 1.42~\AA{} carbon--carbon bonds are resolved. The hexagonal lattice is made commensurate with the square periodic cell by a 3.8\% compression along $y$. The projected potential is assembled atom by atom with \texttt{quscope.ctem.KirklandPotential.calculate\_2d}, pixel-integrated by $4\times$ supersampling and thermally smeared with a Gaussian of width $\sigma = 0.25$~\AA{} (Debye--Waller); the image is returned by \texttt{QuantumCTEMCircuit.simulate(V)}, which builds and executes the circuit of Fig.~\ref{fig:circuit}; and the fidelity is reported by \texttt{QuantumClassicalValidator.compare(V)} against the FFT-based twin. The image shows the classic dark-atom contrast with bright hexagon centers; the two intensities are indistinguishable, and the quantum--classical state fidelity is unity to within double-precision rounding.

\section{Quantum Multislice CTEM}
\label{sec:multislice}

Thick specimens violate the single-phase-grating picture, since the wave must be propagated between successive thin slices of the specimen. In the multislice formalism~\cite{CowleyMoodie1957,Kirkland2020} slice $j$ applies the transmission function $t_j(\mathbf{r}) = e^{i\sigma V_j(\mathbf{r})}$ followed by Fresnel propagation over the slice thickness $\Delta z$,
\begin{equation}
P(k) = e^{-i \pi \lambda \Delta z\, k^{2}},
\label{eq:fresnel}
\end{equation}
applied in momentum space. QuScope's \texttt{QuantumMultisliceCircuit} realizes this by repeating the block
\[
D[e^{i\sigma V_j}] \;\rightarrow\; \mathrm{QFT} \;\rightarrow\; D[P(k)] \;\rightarrow\; \mathrm{QFT}^{\dagger}
\]
once per slice, then applying the objective-lens stage of Eq.~(\ref{eq:pipeline}). All slice potentials enter as diagonal unitaries on the same $n_q$-qubit register, so specimen thickness costs circuit \emph{depth}, not qubits.

\begin{table}[tb]
\centering
\caption{Validation of QuScope v0.2.0 against classical theory and literature values. At 100, 200, and 300~kV, $\lambda = 0.037014/0.025079/0.019687$~\AA{} [Eq.~(\ref{eq:lambda})] and $\sigma = 0.92446/0.72884/0.65262 \times 10^{-3}$~rad\,V$^{-1}$\AA$^{-1}$ [Eq.~(\ref{eq:sigma})]. Fidelity is $|\braket{\psi_q|\psi_c}|^2$ between quantum and classical wavefunctions under strong scattering.}
\label{tab:validation}
\begin{ruledtabular}
\begin{tabular}{@{}p{0.68\linewidth} p{0.28\linewidth}@{}}
Check & Result \\
\hline
$\lambda$ vs.\ literature (100--300~kV) & exact \\
$\sigma$ vs.\ literature (100--300~kV) & $<0.1\%$ deviation \\
CTF $\chi(k)$ vs.\ Eq.~(\ref{eq:chi}) & machine precision \\
Propagator vs.\ Eq.~(\ref{eq:fresnel}) & machine precision \\
Multislice fidelity (SrTiO$_3$, 20 slices) & 1.000000 \\
\multicolumn{2}{@{}p{\linewidth}@{}}{\quad per-depth fidelity, every slice: 1.000000} \\
STEM multislice, 1-slice limit & correlation 1.0000 vs.\ \texttt{run\_stem} \\
\end{tabular}
\end{ruledtabular}
\end{table}

We demonstrate the pipeline on SrTiO$_3$ ($a = 3.905$~\AA) viewed along $[100]$, a standard benchmark structure in quantitative STEM/TEM. Ten unit cells of thickness ($39.05$~\AA) are represented as 20 alternating SrO and TiO$_2$ planes ($\Delta z = a/2 = 1.953$~\AA), each built with \texttt{KirklandPotential} on a $2\times2$-cell field of view ($64\times64$ grid at 0.122~\AA/pixel, 12 qubits), under plane-wave illumination at 200~kV; the exit wave is shown directly, with no objective-lens stage applied ($\Delta\!f = 0$, $C_s = 0$). The scattering is strongly dynamical, with $\sigma V$ reaching $1.7$~rad per slice on the Sr columns, well beyond any weak-phase or kinematic regime. Passing the list of slice potentials to \texttt{QuantumMultisliceCircuit.simulate(potentials)} builds and executes the single 20-slice circuit; \texttt{QuantumClassicalMultisliceValidator} implements the identical slice sequence with FFTs. The quantum and classical exit waves agree to fidelity $|\braket{\psi_q|\psi_c}|^2 = 1.000000$ (Table~\ref{tab:validation}), with residual amplitude differences at the $10^{-10}$ level set by accumulated floating-point rounding across the 80 high-level synthesized gates in the 20-slice circuit (40 diagonal phase gates and 40 QFT/IQFT transforms; Fig.~\ref{fig:multislice}).

Because the quantum state is available after every slice, the framework can also watch the beam \emph{inside} the crystal. Figure~\ref{fig:propagation} shows an aberration-free focused probe (18~mrad semi-angle, $\Delta\!f = 0$, $C_s = 0$, on the same $2\times2$-cell grid) placed on a Sr column and on a Ti--O column, propagated through all 20 slices by executing the slice subcircuits in sequence. Each slice is one phase-grating \texttt{DiagonalGate} followed by the \texttt{QFTGate}-conjugated Fresnel propagator built from \texttt{fresnel\_propagator\_phase}, applied to the state with Qiskit's \texttt{Statevector.evolve}. Each horizontal line of the map is the probe intensity along a lattice row at one depth. The maps show the classic electron-channeling behavior. The incident probe contracts onto the atomic column within the first few unit cells and then breathes periodically along it, with a channeling period that differs between the strong Sr and the weaker Ti--O column. At every one of the 20 depths the circuit-evolved state matches the classical multislice wave to fidelity $1.000000$.

\section{Quantum STEM: Single-Slice and Multislice}
\label{sec:stem}

\subsection{Probe formation and scanning}

In scanning TEM the illumination is a focused, aberrated probe. QuScope forms it in momentum space as $A(k)\,e^{-i\chi(k)}$, where $A$ is the aperture function of semi-angle $\alpha$ (typically 15--25~mrad) and $\chi$ is the probe-forming aberration function of Eq.~(\ref{eq:chi}); the real-space probe at scan position $\mathbf{r}_s$ follows by inverse transform and shift. For each scan position the probe state is amplitude-encoded on the register and propagated through the specimen, using a single phase grating in \texttt{run\_stem} (POA), or the full slice--propagate sequence of Sec.~\ref{sec:multislice} in \texttt{run\_stem\_multislice}. The exit wave is transformed to the detector plane, where its momentum distribution $|\psi(k)|^2$ is integrated over annular masks.

\subsection{Detector implementation}

\begin{figure*}[tb]
\centering
\includegraphics[width=\textwidth]{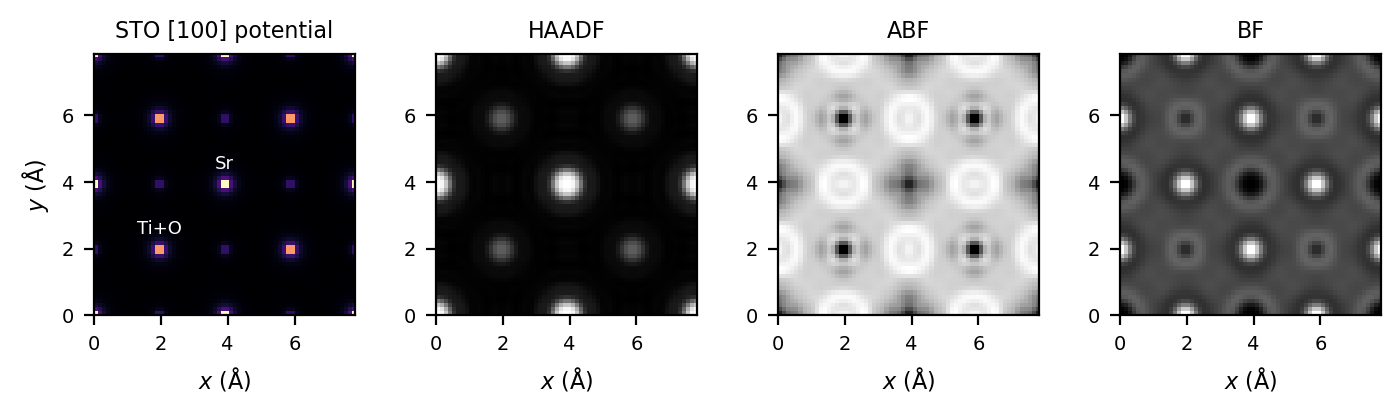}
\caption{Quantum multislice STEM of SrTiO$_3$ $[100]$, ten unit cells thick, at 200~kV (30~mrad convergence semi-angle; one unit cell scanned at full resolution with periodic boundaries, displayed tiled $2\times2$). From left to right, the panels show the projected potential with the Sr and Ti+O columns annotated, the HAADF image dominated by the heavy Sr columns (Z-contrast), the ABF image which is contrast-reversed and sensitive to the lighter columns, and the BF image. Every scan pixel is one complete 20-slice quantum circuit, executed in exact statevector simulation ($1024$ circuits per image).}
\label{fig:stem}
\end{figure*}

QuScope's \texttt{STEMDetectors} implements the standard channels ~\cite{Pennycook1991}, including high-angle annular dark field (HAADF, 60--200~mrad), annular dark field (ADF, 25--60~mrad), annular bright field (ABF, 10--25~mrad), and bright field (BF, $<$10~mrad). Detector angles $\theta$ map to spatial frequencies as $k=\theta/\lambda$; the sampling must satisfy $\theta_{\max}/\lambda < 1/(2\,\Delta x)$ (the grid Nyquist frequency) for the outer detectors to be represented. This is a practical constraint the documentation makes explicit, since an under-sampled detector integrates to zero. At the 0.122~\AA{} sampling of Fig.~\ref{fig:stem} the Nyquist limit corresponds to $\approx103$~mrad, so the HAADF channel effectively collects over 60--103~mrad rather than its nominal 60--200~mrad outer angle; this truncation rescales the absolute HAADF signal but not the Z-contrast between columns.

\begin{figure*}[tb]
\centering
\includegraphics[width=\textwidth]{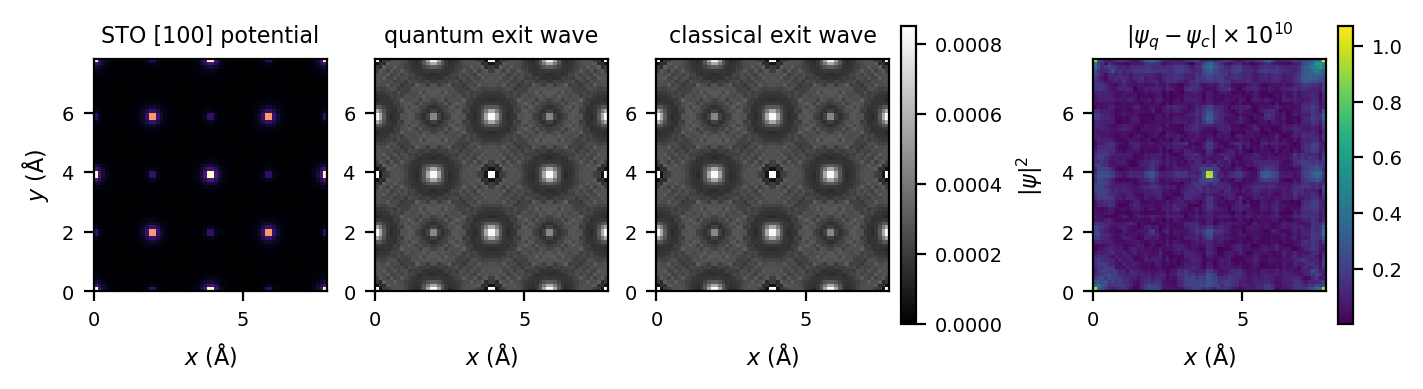}
\caption{Quantum versus classical multislice for SrTiO$_3$ $[100]$, ten unit cells (20 alternating SrO/TiO$_2$ slices) at 200~kV; $2\times2$ unit cells, $64\times64$ grid, 12 qubits. From left to right, the panels show the projected potential, the exit-wave intensity from the single 20-slice quantum circuit showing channeling peaks on the atomic columns, the classical FFT multislice result, and the absolute wavefunction difference (floating-point accumulation scale). Fidelity $=1.000000$. Exact statevector simulation.}
\label{fig:multislice}
\end{figure*}

Figure~\ref{fig:stem} shows quantum multislice STEM of the same SrTiO$_3$ $[100]$ crystal used in Sec.~\ref{sec:multislice}, ten unit cells thick, with an aberration-free 30~mrad probe ($\Delta\!f = 0$, $C_s = 0$) at 200~kV. One unit cell is scanned at full resolution ($32\times32$ grid at 0.122~\AA/pixel, 10 qubits) under periodic boundary conditions, and \emph{every scan pixel is the readout of one complete 20-slice quantum circuit, executed in exact statevector simulation}. Each pixel involves state preparation of the shifted probe, then the multislice circuit assembled by \texttt{build\_probe\_circuit} (twenty phase-grating diagonal unitaries interleaved with QFT-conjugated Fresnel propagators), then the detector-plane transform with the exit intensity integrated over the annular masks of \texttt{STEMDetectors.masks} ($32\times32 = 1024$ such circuits). The images reproduce the textbook contrast of this benchmark structure. HAADF is dominated by the heavy Sr columns ($Z=38$), with weaker Ti+O columns and invisible pure-oxygen columns showing the expected Z-contrast~\cite{Pennycook1991}, while ABF reverses the contrast and picks up the light columns. As a further consistency check, the single-slice limit of \texttt{run\_stem\_multislice} reproduces \texttt{run\_stem} with image correlation $1.0000$ on every channel (Table~\ref{tab:validation}).
\begin{figure}[tb]
\centering
\includegraphics[width=0.9\columnwidth]{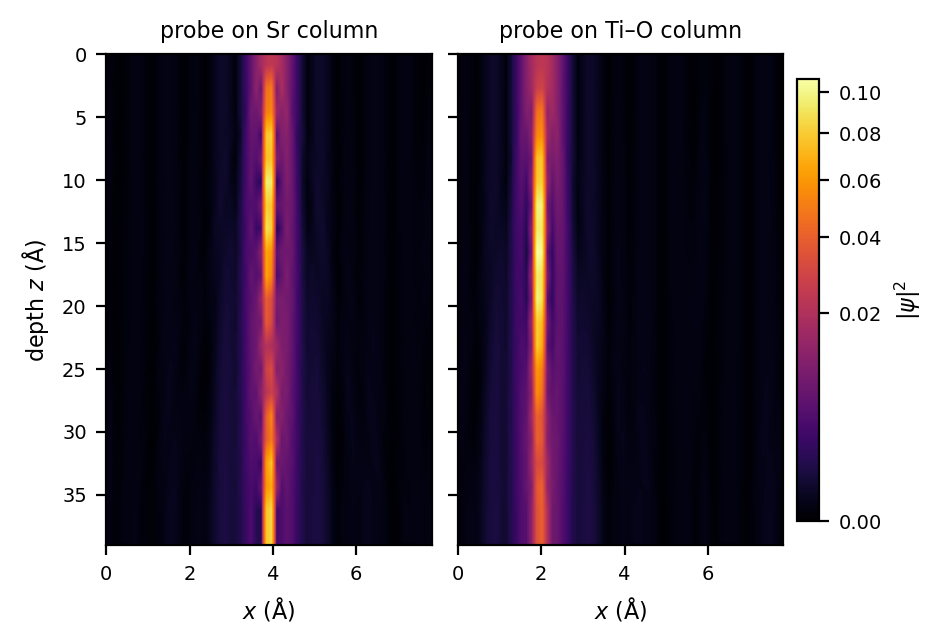}
\caption{An 18-mrad probe channeling through SrTiO$_3$ $[100]$ (ten unit cells), computed by executing the 20 slice subcircuits in sequence. The maps show probe intensity $|\psi(x, z)|^2$ along the lattice row containing the column, versus depth $z$ (beam enters at the top; power-law color scale). The left panel shows the probe on a Sr column and the right panel shows the probe on a Ti--O column. The probe contracts onto the column and oscillates with a column-dependent channeling period. Quantum and classical state fidelity is $1.000000$ at every depth. Exact statevector simulation.}
\label{fig:propagation}
\end{figure}

The package's default \texttt{run\_stem\_multislice} scan loop applies the diagonal gates as their exact statevector action (an elementwise phase multiplication, mathematically identical to the synthesized gate), while executing the QFTs as circuits. The fully gate-level path is exposed through \texttt{build\_probe\_circuit} and is what all figures in this paper use.

\section{Validation Against Classical Theory}
\label{sec:validation}

Table~\ref{tab:validation} summarizes the validation suite. Physical constants are checked against standard reference values~\cite{Kirkland2020}; circuit stages are checked against their closed-form definitions; and full pipelines are checked against the classical twin implementations. All checks are reproducible from the repository, and the pipeline-level comparisons run in continuous integration on every commit.

Two remarks on methodology are worth noting. Fidelity is evaluated under \emph{strong} scattering (the SrTiO$_3$ benchmark reaches $\sigma V = 1.7$~rad per slice on the Sr columns, sustained over 20 slices). Weakly scattering test potentials leave the wave nearly planar and would mask propagator errors, so a fidelity of unity is only meaningful when the dynamics are nontrivial. In addition, the constants in Table~\ref{tab:validation} are validated at the level of the public functions (\texttt{relativistic\_wavelength}, \texttt{interaction\_constant}) rather than only implicitly through pipeline tests, so a regression in either is caught directly.

\section{Quantum Resource Analysis}
\label{sec:resources}

Table~\ref{tab:resources} reports the resources of the actual circuits, generated programmatically from the released code. Each pipeline is built at $8\times8$ (6 qubits) and $16\times16$ (8 qubits), and transpiled to the hardware-native basis $\{R_z, \sqrt{X}, X, \mathrm{CX}\}$ at Qiskit optimization level 1. Both the logical (high-level) operation count and the transpiled counts are shown, and the generating script ships with the paper source so the table can be regenerated on any code version.

\begin{table*}[htb]
    \centering
    \caption{Quantum resources for QuScope v0.2.0 algorithms, measured from the actual circuits. Logical resources count high-level operations (DiagonalGate, QFTGate, state preparation); transpiled resources use the $\{R_z, \sqrt{X}, X, \mathrm{CX}\}$ basis at optimization level 1.}
    \label{tab:resources}
    \begin{ruledtabular}
    \begin{tabular}{llccccc}
    Algorithm & Grid & Qubits & Logical ops & Logical depth & CX & Transpiled depth \\
    \hline
    CTEM pipeline (WPOA+CTF) & 8x8 & 6 & 12 & 5 & 160 & 266 \\
    Multislice CTEM (4 slices) & 8x8 & 6 & 30 & 17 & 640 & 894 \\
    STEM probe circuit (3 slices) & 8x8 & 6 & 13 & 9 & 382 & 534 \\
    STEM probe circuit (single slice) & 8x8 & 6 & 1 & 1 & 62 & 118 \\
    CTEM pipeline (WPOA+CTF) & 16x16 & 8 & 14 & 5 & 580 & 1016 \\
    Multislice CTEM (4 slices) & 16x16 & 8 & 32 & 17 & 2320 & 3390 \\
    STEM probe circuit (3 slices) & 16x16 & 8 & 13 & 9 & 1414 & 2066 \\
    STEM probe circuit (single slice) & 16x16 & 8 & 1 & 1 & 254 & 498 \\
    \end{tabular}
    \end{ruledtabular}
\end{table*}

Two features of Table~\ref{tab:resources} carry the physics of Sec.~\ref{sec:bottleneck}. The logical structure is nearly flat, since the CTEM pipeline is a handful of high-level operations regardless of grid size. The transpiled counts, by contrast, roughly quadruple per grid doubling. This is the $\mathcal{O}(2^{n_q})$ synthesis cost of arbitrary diagonal unitaries~\cite{Shende2006,Welch2014} made concrete. Depth is also dominated by the diagonal gates rather than the QFTs. At $16\times16$ the two-dimensional QFT contributes tens of gates while each arbitrary diagonal contributes hundreds of CX gates. Extrapolating, a $256\times256$ simulation (16 qubits) requires of order $10^5$ two-qubit gates per diagonal. This is beyond NISQ error budgets but firmly within early fault-tolerant regimes, and it is dramatically reducible for potentials with structure (Sec.~\ref{sec:outlook}).

We deliberately refrain from reporting hardware execution times or fidelities, since such figures become outdated rapidly and depend on device-specific calibration. Instead, we give illustrative estimates using representative, literature-typical error rates (below) to contextualize the resource counts in Table~\ref{tab:resources}. For up-to-date, device-specific numbers, QuScope ships transpilation and device-profile utilities so users can generate current estimates for the backend they run on.

\subsection{Connectivity and execution on real devices}
\label{sec:hardware}

QuScope connects to real hardware through its backend layer. The \texttt{quantum\_ctem.backends} package wraps Aer simulators and IBM Quantum systems behind a common interface, and the \texttt{ibm\_config} module handles the session plumbing (\texttt{load\_ibm\_credentials} from an environment token, \texttt{get\_ibm\_service}, \texttt{list\_available\_backends}, \texttt{validate\_ibm\_access}). Any circuit shown in this paper can be retargeted to a device with Qiskit's transpiler or the package's \texttt{HardwareTranspiler}; the basis $\{R_z, \sqrt{X}, X, \mathrm{CX}\}$ used in Table~\ref{tab:resources} is the IBM native gate set, so those counts translate directly.

Two effects separate Table~\ref{tab:resources} from what a physical device executes. The first is \emph{connectivity}. The counts in Table~\ref{tab:resources} assume all-to-all coupling, while superconducting devices provide a sparse (heavy-hexagonal, degree $\leq 3$) qubit graph. The QFT's long-range controlled phases and the entangling ladders of diagonal-gate synthesis then require SWAP routing, which typically inflates two-qubit counts by a further factor of $2$--$3\times$ on heavy-hex topologies. The \texttt{ibm\_hardware\_validation} module (\texttt{IBMDeviceProfile}, \texttt{estimate\_fidelity}, \texttt{validate\_ibm\_deployment}) folds a device's reported connectivity, gate errors, and coherence times into a per-circuit success estimate for exactly this purpose. The second effect is \emph{noise}. At a representative two-qubit error of $10^{-3}$, an $8\times8$ CTEM pipeline ($\sim$600 CX before routing) retains on the order of $50\%$ circuit fidelity, demonstrable with readout mitigation and zero-noise extrapolation, while the deeper multislice circuits fall below $10\%$ and the $16$-qubit extrapolations of Sec.~\ref{sec:resources} are out of reach for any error-mitigation strategy.

That boundary is where quantum error correction enters. Encoding each register qubit in a distance-$d$ surface code costs roughly $2d^2$ physical qubits per logical qubit ($\sim10^3$ at current physical error rates for logical error rates near $10^{-9}$). The full $256\times256$ pipeline, with 16 logical qubits and $\sim10^5$ two-qubit gates per diagonal, therefore maps to a few times $10^4$ physical qubits with comfortable logical depth. This is a workload for early fault-tolerant machines rather than present NISQ processors~\cite{Preskill2018}. QuScope's role in that transition is to keep the resource accounting concrete. Because every pipeline is a real circuit, its logical-qubit count, T/rotation budget, and depth can be re-costed against any error-correction scheme without re-deriving the physics.

\section{Software Engineering and Availability}
\label{sec:software}

QuScope v0.2.0 is installable from PyPI (\texttt{pip install quscope}) and developed openly at \url{https://github.com/QuScope/QuScope}, with documentation at \url{https://quscope.readthedocs.io}. The package requires Python $\geq 3.9$ and Qiskit $\geq 2.0$, uses NumPy/SciPy/Matplotlib~\cite{harris2020,virtanen2020,Hunter2007} for classical numerics and visualization, and is MIT-licensed.

Quality controls reflect the validation-first design. The test suite (250+ tests) runs on Python 3.9--3.12 in continuous integration and gates every merge; pipeline-level quantum--classical validators are part of that suite; data files (Kirkland scattering-factor tables) ship inside the package; and the demonstration notebooks are executable documentation, rendered in the online gallery. Release versions follow semantic versioning, with the public API of the four pipelines stable within the 0.2 series.

\section{Limitations and Outlook}
\label{sec:outlook}

\emph{Diagonal synthesis.} As quantified in Sec.~\ref{sec:resources}, loading an arbitrary potential costs $\mathcal{O}(2^{n_q})$ gates, which cancels the exponential state compression in the general case. The escape route is structure. Crystalline potentials are sparse in Fourier space, and a diagonal whose phase profile has $s$ significant Fourier components admits circuits polynomial in $s$~\cite{Welch2014}. Exploiting crystal periodicity in the diagonal synthesis is the single highest-leverage item on the roadmap.

\emph{Readout.} Statevector simulation returns the full wavefunction, but hardware returns samples. Reconstructing a full image from samples costs $\mathcal{O}(N^2)$ shots at fixed precision per pixel, no better than classical, so the near-term-sensible outputs are \emph{low-dimensional observables}. STEM detector signals are integrals of $|\psi(k)|^2$ over annular masks, exactly the kind of quantity for which amplitude-estimation techniques offer quadratic sample advantages. This is why the STEM geometry, not full-image CTEM, is the most natural candidate for early hardware utility.

\emph{Statevector bounds.} Exact simulation is memory-bound near $n_q \approx 30$ on workstations; QuScope's practical envelope with validation overhead is $n_q \leq 16$ ($256\times256$). This bounds present-day demonstrations, not the formalism.

\emph{Roadmap.} Under active development on the project's \texttt{dev} branch are quantum electron-diffraction modes (selected-area and convergent-beam geometries), thermal-diffuse-scattering channels for frozen-phonon physics (a Debye-Waller amplitude-damping channel implemented with one ancilla and uniformly-controlled rotations, a phonon-register superposition method, and an open-system Kraus/Lindblad multislice), and a Bloch-wave eigensolver that extracts Bethe excitation errors by quantum phase estimation. Each will ship with the same classical-twin validation discipline as the four pipelines presented here.

\section{Data and Code Availability}
\label{sec:availability}

QuScope is open source under the MIT license. The source code is available at \url{https://github.com/QuScope/QuScope} (release v0.2.0), the package is distributed at \url{https://pypi.org/project/quscope/}, and documentation and the notebook gallery are hosted at \url{https://quscope.readthedocs.io}. The scripts generating Table~\ref{tab:resources} and Figs.~\ref{fig:ctem}--\ref{fig:stem} are included in the examples repository (\url{https://github.com/QuScope/examples-applications}).

\begin{acknowledgments}
This research was supported in part through the computational resources by the Quest high performance computing facility at Northwestern University, jointly supported by the Office of the Provost, Office for Research, and Northwestern University Information Technology. We acknowledge the use of IBM Quantum services for this work~\cite{qiskit2024}. The views expressed are those of the authors, and do not reflect the official policy or position of IBM or the IBM Quantum team. SDL thanks the support of the MRSEC program (NSF DMR-2308691) at the Northwestern Materials Research Center. SDL thanks the Materials Initiative for Comprehensive Research Opportunity (MICRO) program which benefited greatly from administrative and research support at MIT and Northwestern and from an unrestricted gift from the 3M Foundation (3M STEM and Skilled Trades Program).
\end{acknowledgments}

%

\end{document}